\documentclass[journal,12pt,onecolumn,draftclsnofoot,]{IEEEtran}

\ifCLASSINFOpdf
\else
   \usepackage[dvips]{graphicx}
\fi
\usepackage{url}

\usepackage[noadjust]{cite}
\usepackage{amsmath,amssymb,amsfonts}
\usepackage{algorithmic}
\usepackage{graphicx}
\usepackage{textcomp}
\usepackage{xcolor}
\usepackage{hhline}
\usepackage{multirow}
\usepackage{amsmath}
\usepackage{graphicx}
\usepackage[british]{babel}
\usepackage{amsfonts}
\usepackage{amssymb}
\usepackage{graphicx}
\usepackage[export]{adjustbox}
\usepackage{mathtools}
\usepackage{siunitx}
\usepackage{makecell}
\usepackage[pagebackref=true,colorlinks=false,breaklinks=true,letterpaper=true,bookmarks=false]{hyperref} 
\usepackage{algorithmic}
\usepackage{ctable} 
\usepackage{gensymb}
\usepackage{array}
\usepackage{threeparttable}
\usepackage{balance}

\newcommand{\etal}{\textit{et al}.~}
\newcommand{\ie}{\textit{i}.\textit{e}.~}

\newcommand{\disp}{\mathrm{disp}}
\newcommand{\range}{\mathrm{range}}
\newcommand{\shear}{\mathrm{shear}}
\newcommand{\crop}{\mathrm{crop}}

\newcommand{\cyc}{\mathrm{cyc}}

\newcommand\norm[1]{\left\lVert#1\right\rVert}

\newcommand{\figref}[1]{Fig.\,\ref{#1}}
\newcommand{\tabref}[1]{Table\,\ref{#1}}
\newcommand{\secref}[1]{Section\,\ref{#1}}

\renewcommand\thetable{\Roman{table}}

\newcommand{\RNum}[1]{\uppercase\expandafter{\romannumeral #1\relax}}

\def\wrt{\textit{w.r.t.~}}

\begin{document}

\title{Self-Supervised Light Field Reconstruction Using Shearlet Transform and \\ Cycle Consistency}

\author{
Yuan Gao, 
Robert Bregovi\'{c}, \IEEEmembership{Member, IEEE}, 
and Atanas Gotchev, \IEEEmembership{Member, IEEE}

%\thanks{This paragraph of the first footnote will contain the date on which you submitted your paper for review. It will also contain support information, including sponsor and financial support acknowledgment. For example, ``This work was supported in part by the U.S. Department of Commerce under Grant BS123456.'' }
\thanks{Y. Gao, R. Bregovi\'{c} and A. Gotchev are with the Faculty of Information Technology and Communication Sciences (ITC), Tampere University, FI-33720 Tampere, Finland 
(e-mail: \{\href{mailto:yuan.gao@tuni.fi}{yuan.gao}, \href{mailto:robert.bregovic@tuni.fi}{robert.bregovic}, \href{mailto:atanas.gotchev@tuni.fi}{atanas.gotchev}\}@tuni.fi).}
%\thanks{S. B. Author, Jr., was with Rice Univeersity, Houston, TX 77005 USA. He is now with the Department of Physics, Colorado State University, Fort Collins, CO 80523 USA (e-mail: author@lamar.colostate.edu).}
}

\markboth{Journal of \LaTeX\ Class Files, Vol. 14, No. 8, August 2015}
{Shell \MakeLowercase{\textit{et al.}}: Bare Demo of IEEEtran.cls for IEEE Journals}
\maketitle

\begin{abstract}
The image-based rendering approach using Shearlet Transform (ST) is one of the state-of-the-art Densely-Sampled Light Field (DSLF) reconstruction methods.
It reconstructs Epipolar-Plane Images (EPIs) in image domain via an iterative regularization algorithm restoring their coefficients in shearlet domain.
Consequently, the ST method tends to be slow because of the time spent on domain transformations for dozens of iterations.
To overcome this limitation, this letter proposes a novel self-supervised DSLF reconstruction method, CycleST, which applies ST and cycle consistency to DSLF reconstruction.   
Specifically, CycleST is composed of an encoder-decoder network and a residual learning strategy that restore the shearlet coefficients of densely-sampled EPIs using EPI reconstruction and cycle consistency losses.
Besides, CycleST is a self-supervised approach that can be trained solely on Sparsely-Sampled Light Fields  (SSLFs) with small disparity ranges ($\leqslant$ 8 pixels).
%In addition, theoretically,
%the different DSLFs reconstructed from the same SSLF by the same method should have identical results for the same angular positions, 
%which is fully leveraged by CycleST as circle consistency loss. 
%Moreover, CycleST is trained only on SSLFs with small disparity ranges ($\leqslant$ 8 pixels).
Experimental results of DSLF reconstruction on SSLFs with large disparity ranges (16\,-\,32 pixels) from two challenging real-world light field datasets demonstrate the effectiveness and efficiency of the proposed CycleST method.
Furthermore, CycleST achieves $\thicksim$\,9x speedup over ST, at least. 
\end{abstract}


\IEEEpeerreviewmaketitle

\vspace{-.5em}
\section{Introduction}

\IEEEPARstart{D}{ensely}-Sampled Light Field (DSLF) is a discrete 4D representation for the light rays from the scene encoded by two parallel planes, namely image plane and camera plane,
where the disparity ranges between neighboring views are less or equal to one pixel. 
The DSLF has a wide range of applications, such as synthetic aperture imaging, depth estimation, segmentation and visual odometry 
\cite{wu2017light}.
Besides, the DSLF-based contents can be rendered on  
VR \cite{yu2017light},
3DTV \cite{smolic20113d} and
holographic \cite{agocs2006large} systems.
A DSLF in real-world environment is extremely difficult to capture, 
due to the hardware limitations of the modern light field acquisition systems that can in most cases only capture Sparsely-Sampled Light Fields (SSLFs), 
where the disparity range between views is more than one pixel. 
Therefore, real-world DSLFs are typically reconstructed from real-world SSLFs using computational imaging approaches.
\par
\noindent\textbf{\emph{Related work}}. 
Video frame interpolation methods can be adapted to solve the DSLF reconstruction problem because a 3D SSLF can be treated as a virtual video sequence. 
Niklaus \etal have proposed Separable Convolution (SepConv), a learning-based video frame synthesis method using spatially adaptive kernels
\cite{niklaus2017iccv}.
Bao \etal have proposed a Depth-Aware video frame INterpolation (DAIN) algorithm that leverages optical flow, local interpolation kernels, depth maps and contextual features
\cite{bao2019depth}.  
Xu \etal have improved the linear models adopted in Liner Video Interpolation (LVI) methods
\cite{jiang2018super,liu2017video}
and proposed the Quadratic Video Interpolation (QVI) approach considering the acceleration information in videos
\cite{xu2019quadratic, li2019quadratic,nah2019aim}.
Gao and Koch were the first to extend SepConv for DSLF reconstruction by proposing Parallax-Interpolation Adaptive Separable Convolution (PIASC)
\cite{gao2018icmew}.
More recently, Gao \etal have developed a learning-based DSLF reconstruction method, \ie Deep Residual Shearlet Transform (DRST)
\cite{gao2019tcsvt}, 
for coefficient restoration in the shearlet domain of densely-sampled Epipolar-Plane Images (EPIs)
based on the conventional image-based rendering techniques 
\cite{shum2008image,shum2003survey}
using Shearlet Transform (ST) \cite{vagharshakyan2018light,vagharshakyan2017accelerated,vagharshakyan2015image}.

\par
As for DSLF reconstruction on $\emph{large-disparity-range}$ (16\,-\,32 pixels) SSLFs  of complex scenes, 
the aforementioned state-of-the-art video frame interpolation methods tend to fail.
To tackle this problem, a novel self-supervised DSLF reconstruction method, CycleST, is proposed by leveraging shearlet transform and cycle consistency
\cite{zhu2017unpaired}, 
also used by recent video frame interpolation methods
\cite{reda2019unsupervised,liu2019deep}. 
In particular, since several DSLFs with different angular resolutions can be reconstructed from the same input SSLF,
the cycle consistency is the technique that
guarantees these DSLFs have similar reconstruction results \wrt the same angular positions. 
%Assume two DSLFs are reconstructed from this SSLF with different numbers of synthesized views, \ie ($\tau-1$) and ($2\tau-1$), between adjacent views of the input SSLF.
%The synthesized ($\tau-1$) images should be identical to those in the ($2\tau-1$) images with the same angular positions if the DSLF reconstruction method is robust.
%This constraint is referred to as cycle consistency

\par
We summarize the main contributions of this letter as below:
\begin{itemize}
\item The proposed CycleST fully leverages the deep convolutional network with cycle consistency loss in shearlet domain to perform EPI inpainting in image domain. 
\item CycleST is fully self-supervised and trained solely on synthetic SSLFs with \emph{small disparity ranges} ($\leqslant$ 8 pixels);
\item Experimental results on challenging real-world SSLFs with large disparity ranges (16\,-\,32 pixels) demonstrate the superiority of CycleST over ST for DSLF reconstruction in terms of accuracy and efficiency ($>$ 8.9x speedup). 
\end{itemize}

\vspace{-.5em}
\section{Methodology}
%\vspace{-.5em}
%The proposed CycleST approach will be presented after a brief introduction to the DSLF reconstruction problem in the following section.
\subsection{Preliminaries} \label{sec:preliminary}
\subsubsection{Symbols and notations} \label{sec:symbols}
The symbols and notations used by this letter are elaborated in
\tabref{tab:symbol}.
As can be seen in the table, the target DSLF $\mathcal{D}$ to be reconstructed from the input SSLF $\mathcal{S}$ has the same spatial resolution ($m\times l$) but different angular resolutions.
Specifically, on the one hand, the angular resolution of $\mathcal{D}$, \ie $\dot{n}$, is dependent on that of $\mathcal{S}$, \ie $n$, and the sampling interval $\tau$ specified by the user such that $\dot{n}=\big((n-1)\tau+1\big)$.
On the other hand, $\tau$ relies on the disparity range $d_{\range}^{\mathcal{S}}$ of $\mathcal{S}$, \ie $\tau \geqslant d_{\range}^{\mathcal{S}}$.
Since this paper aims to enhance the angular resolution of any input $\mathcal{S}$ with \emph{large disparity ranges} ($16 < d_{\range}^{\mathcal{S}} \leqslant 32$ pixels), we fix $\tau$ to 32.
Both $\mathcal{S}$ and $\mathcal{D}$ can be regarded as sets of EPIs, \ie $\mathcal{S}=\{\varepsilon_i | 1 \leqslant i \leqslant l\}$ and $\mathcal{D}=\{\zeta_i | 1 \leqslant i \leqslant l\}$.
It is obvious that $\varepsilon_i$ and $\zeta_i$ have different heights because of the different angular resolutions.
% training dataset 
%In order to apply the learning-based methods to solve the DSLF reconstruction problem, 
%one needs to prepare the training data of $\mathcal{D}$, 
%which can be typically generated by means of modern 3D graphics software.
%However, the rendering of synthetic DSLFs is a tedious process, 
%resulting in few DSLF datasets released by the light field research community.
%Most of the available community-supported light field datasets are SSLFs with \emph{small disparity ranges} ($\leqslant$\,8 pixels), 
%which are used as the training data of this letter.
Due to the lack of DSLF datasets, in this letter for the CycleST network training we utilize the available community-supported SSLF datasets with \emph{small disparity ranges} ($\leqslant$\,8 pixels).
%because the time costs of them are much less than those of DSLFs.  
Let $\mathcal{S}'=\{\epsilon_i | 1 \leqslant i \leqslant l\}$ denote one of these training SSLFs, 
where $d_{\range}^{\mathcal{S}'} \leqslant 8$ pixels and $\epsilon_i \in \mathbb{R}^{m \times n' \times 3}$.
The sampling interval from $\mathcal{D}$ to $\mathcal{S}'$ is represented by
%Note that, 
%in order to sample a $\mathcal{S}$ from a $\mathcal{S}'$, 
%similar to $\tau$,
$\tau'$, where $\tau' \geqslant d_{\range}^{\mathcal{S}'}$. %, where $\tau'$ is the sampling interval 
%Similar to setting $\tau$, 
Here, we fix $\tau'$ to 8.
%$(n'-1)\% (n-1) = 0$, where `$\%$' is the modulo operator,
This letter studies how to reconstruct \emph{densely-sampled} $\mathcal{D}$ from \emph{large-disparity-range} $\mathcal{S}$ using a deep neural network trained solely on \emph{small-disparity-range} $\mathcal{S}'$ in a self-supervised fashion.

\begin{table}
\begin{center}
\caption{Symbols and notations.}
\vspace{-.8em}
\label{tab:symbol}
\small
\setlength{\tabcolsep}{3pt}
%\begin{tabular}{|p{25pt}|p{75pt}|p{110pt}|}
\scalebox{.85}{
\begin{tabular}{@{}lll@{}}
%\hline
\specialrule{.15em}{.0em}{.2em}
Symbol& 
Name& 
Description \\
\midrule
$m \times l$ & Spatial resolution of $\mathcal{S}$, $\mathcal{S}'$ and $\mathcal{D}$  & width $\times$ height\\
$n$, $n'$, $\dot{n}$ & Angular resolutions of $\mathcal{S}$, $\mathcal{S}'$ and $\mathcal{D}$ & \\
$\varepsilon_i$ & Sparsely-sampled EPI (SSEPI) & $\varepsilon_i \in \mathbb{R}^{m \times n \times 3}$ \\
$\epsilon_i$ & SSEPI (for training) & $\epsilon_i \in \mathbb{R}^{m \times n' \times 3}$ \\
$\zeta_i$ & Densely-sampled EPI & $\zeta_i \in \mathbb{R}^{m \times \dot{n} \times 3}$\\
\multirow{2}{*}{$\mathcal{S}$} & Large-disparity-range SSLF & \multirow{2}{*}{$\mathcal{S}=\{\varepsilon_i | 1 \leqslant i \leqslant l\}$} \\
& ($16 < \disp(\mathcal{S}) \leqslant 32$ pixels) & \\
\multirow{2}{*}{$\mathcal{S}'$} & Small-disparity-range SSLF (for training) & \multirow{2}{*}{$\mathcal{S}'=\{\epsilon_i | 1 \leqslant i \leqslant l\}$} \\
& ($1 < \disp(\mathcal{S}') \leqslant 8$ pixels) & \\
\multirow{2}{*}{$\mathcal{D}$} & Target DSLF to be reconstructed from $\mathcal{S}$ & \multirow{2}{*}{$\mathcal{D}=\{\zeta_i | 1 \leqslant i \leqslant l\}$} \\
& ($\disp(\mathcal{D}) \leqslant 1$ pixel) & \\
$d_{\min}^{\mathcal{S}}$ & Minimum disparity of $\mathcal{S}$ & $d_{\min}^{\mathcal{S}} = \mathrm{min}\big(\disp(\mathcal{S})\big)$ \\
$d_{\max}^{\mathcal{S}}$ & Maximum disparity of $\mathcal{S}$ & $d_{\max}^{\mathcal{S}} = \mathrm{max}\big(\disp(\mathcal{S})\big)$ \\
$d_{\range}^{\mathcal{S}}$ & Disparity range of $\mathcal{S}$ & $d_{\range}^{\mathcal{S}} = \big(d_{\max}^{\mathcal{S}} - d_{\min}^{\mathcal{S}}\big)$ \\
$\tau$ & Sampling interval ($\mathcal{D} \rightarrow \mathcal{S}$) & $\tau = \frac{\dot{n}-1}{n-1}=32$ \\ 
$\tau'$ & Sampling interval ($\mathcal{D} \rightarrow \mathcal{S}'$) & $\tau' = \frac{\dot{n}-1}{n'-1}=8$ \\ 
\specialrule{.15em}{.2em}{.0em}
\end{tabular}
}
\end{center}
\vspace{-.2em}
\end{table}

\subsubsection{Shearlet Transform (ST)}	\label{sec:st}
ST is originally proposed in 
\cite{shearlab2016,kutyniok2012shearlab,kutyniok2012shearlets} and extended by Vagharshakyan \etal for DSLF reconstruction 
\cite{vagharshakyan2018light,vagharshakyan2017accelerated,vagharshakyan2015image},  
where an elaborately-tailored shearlet system with $\xi$ scales is developed for the angular resolution enhancement of any input $\mathcal{S}$ with two requirements:
(i) $d_{\min}^{\mathcal{S}} \geqslant 0$ and (ii) $d_{\max}^{\mathcal{S}} \leqslant \tau$.
The number of scales, \ie $\xi$, is determined by the sampling interval $\tau$ as $\xi={\lceil\log_2{\tau\rceil}}$.
The constructed $\xi$-scale shearlet system is
used by shearlet analysis transform $\mathcal{SH} : \mathbb{R}^{\gamma \times \gamma} \rightarrow \mathbb{R}^{\gamma \times \gamma \times \eta}$ and shearlet synthesis transform $\mathcal{SH}^* : \mathbb{R}^{\gamma \times \gamma \times \eta} \rightarrow \mathbb{R}^{\gamma \times \gamma}$,
where $\gamma \times \gamma$ represents the size of a shearlet filter and $\eta = (2^{\xi+1}+\xi-1)$ denotes the number of shearlets.
For $\tau=32$,
$\xi=5$ and $\eta=68$.
Moreover, as suggested in \cite{vagharshakyan2018light}, 
for $\tau=32$, a good choice for $\gamma$ is $\gamma = 255$. 
%which is also adopted in this letter.

\vspace{-.5em}
\subsection{CycleST} 
To resolve the challenging DSLF reconstruction problem for \emph{large-disparity-range} SSLFs using \emph{small-disparity-range} SSLF data only, 
we propose a novel self-supervised method that leverages a residual learning-based convolutional network with 
EPI reconstruction and cycle consistency losses to restore EPI coefficients in shearlet domain.  
The proposed approach is referred to as CycleST and
consists of five steps, 
namely (i) pre-shearing, (ii) random cropping, (iii) remapping, (iv) sparse regularization and (v) post-shearing.
It is worth remarking that steps (i), (ii), (iii) and (iv) belong to the self-supervised training part of CycleST and steps (i), (iii), (iv) and (v) constitute the prediction part of CycleST. 
The details of these five steps are described as following. 
%Explain here below is for training.

\begin{figure}[t]
\begin{minipage}[m]{1.\linewidth}
  \centering
  \centerline{\includegraphics[width=1.\textwidth]{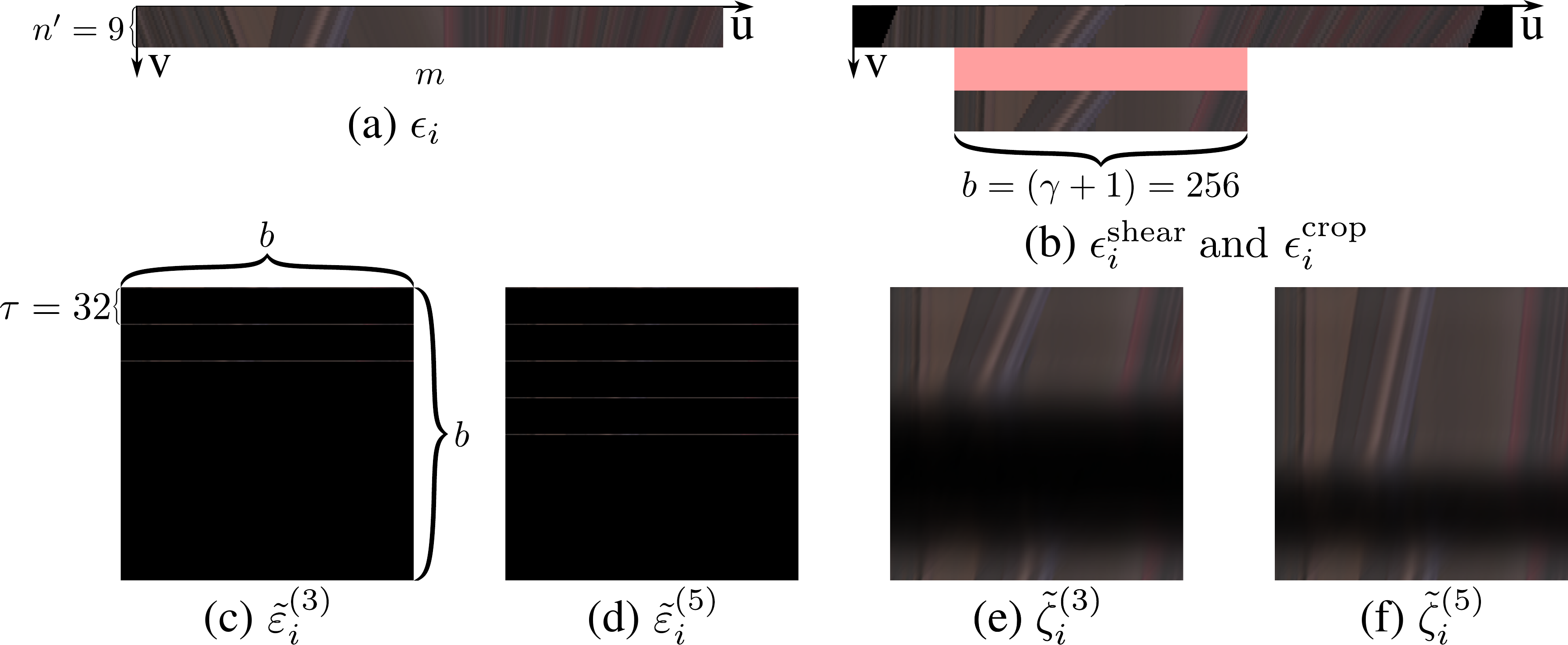}}
\end{minipage}
\vspace{-.8em}
\caption{Illustration of the preparation of training data for CycleST. An EPI $\epsilon_i$ of the training SSLF $\mathcal{S}'$ is presented in (a). In (b), the top row shows $\epsilon^{\shear}_i$, \ie the result of the pre-shearing operation on $\epsilon_i$, and the bottom row shows $\epsilon^{\crop}_i$, which is the  result of the random cropping step on $\epsilon^{\shear}_i$. 
The remapped EPIs $\tilde{\varepsilon}_i^{(3)}$ and $\tilde{\varepsilon}_i^{(5)}$ from $\epsilon^{\crop}_i$ are exhibited in (c) and (d), respectively. 
The corresponding reconstruction results of the sparse regularization step, \ie $\tilde{\zeta}_i^{(3)}$ and $\tilde{\zeta}_i^{(5)}$, are displayed in (e) and (f), respectively.
}
\vspace{.4em}
\label{fig:epi_all}
\end{figure}

\subsubsection{Pre-shearing}	\label{sec:shear}
To satisfy the two requirements of the elaborately-tailored shearlet system explained in \secref{sec:st}, 
a pre-shearing operation is designed to change the disparities of the input training SSLF $\mathcal{S}'$ using a shearing parameter $\rho^{\mathcal{S}'}$, 
where $\Big(d_{\min}^{\mathcal{S}'} - \big(\tau' - d_{\range}^{\mathcal{S}'}\big) \Big) \leqslant \rho^{\mathcal{S}'} \leqslant d_{\min}^{\mathcal{S}'}$.
To be precise, each row $v$ of $\epsilon_i$ is sheared by $(n' - v)\rho^{\mathcal{S}'}$ pixels, where $1 \leqslant v \leqslant n'$.
One of the EPIs of the training SSLF $\mathcal{S}'$, \ie $\epsilon_i$, is displayed in \figref{fig:epi_all}\,(a).
The sheared $\epsilon_i$ is represented by $\epsilon^{\shear}_i$ as illustrated in \figref{fig:epi_all}\,(b).
%The training data are the SSEPIs of $\mathcal{S}'$, 
%one of which is illustrated in \figref{fig:epi_all}\,(a).
%$\epsilon_i^{\shear}=\shear(\epsilon, d)$

\subsubsection{Random cropping}
To augment the number of training samples, a random cropping operation is leveraged to randomly cut an EPI $\epsilon_i^{\crop}$ from the above generated $\epsilon^{\shear}_i$ with a smaller width $b=(\gamma+1)=256$ pixels. 
Note that this operation does not crop any black border region of $\epsilon^{\shear}_i$.
An example of the random cropping results is shown in \figref{fig:epi_all}\,(b).
%$\epsilon_i^{\crop}=\crop\big(\epsilon_i^{\shear}, b \big)$

\subsubsection{Remapping}
To achieve self-supervision for CycleST, 
the rows of the cropped EPI $\epsilon_i^{\crop}$ are rearranged with zero-padding between neighboring rows, 
producing EPIs $\tilde{\varepsilon}_i^{(t)}$, \ie 
\begin{IEEEeqnarray}{c}	
\tilde{\varepsilon}_i^{(t)}\Big(1:\tau:\big((t-1)\tau  + 1\big)\Big) = \epsilon_i^{\crop} \Big(1: \tfrac{n'-1}{t-1}: n'\Big)\,,
\end{IEEEeqnarray}
where $\tilde{\varepsilon}_i^{(t)} \in \mathbb{R}^{b \times b \times 3}$.
As shown in \figref{fig:epi_all}\,(c) and (d),  two different EPIs, \ie $\tilde{\varepsilon}_i^{(3)}$ and $\tilde{\varepsilon}_i^{(5)}$, are generated for each $\epsilon_i^{\crop}$, 
so that the cycle consistency information from them can be leveraged in the next sparse regularization step.

\begin{figure}[t]
\begin{minipage}[m]{1.\linewidth}
  \centering
  \centerline{\includegraphics[width=1.\textwidth]{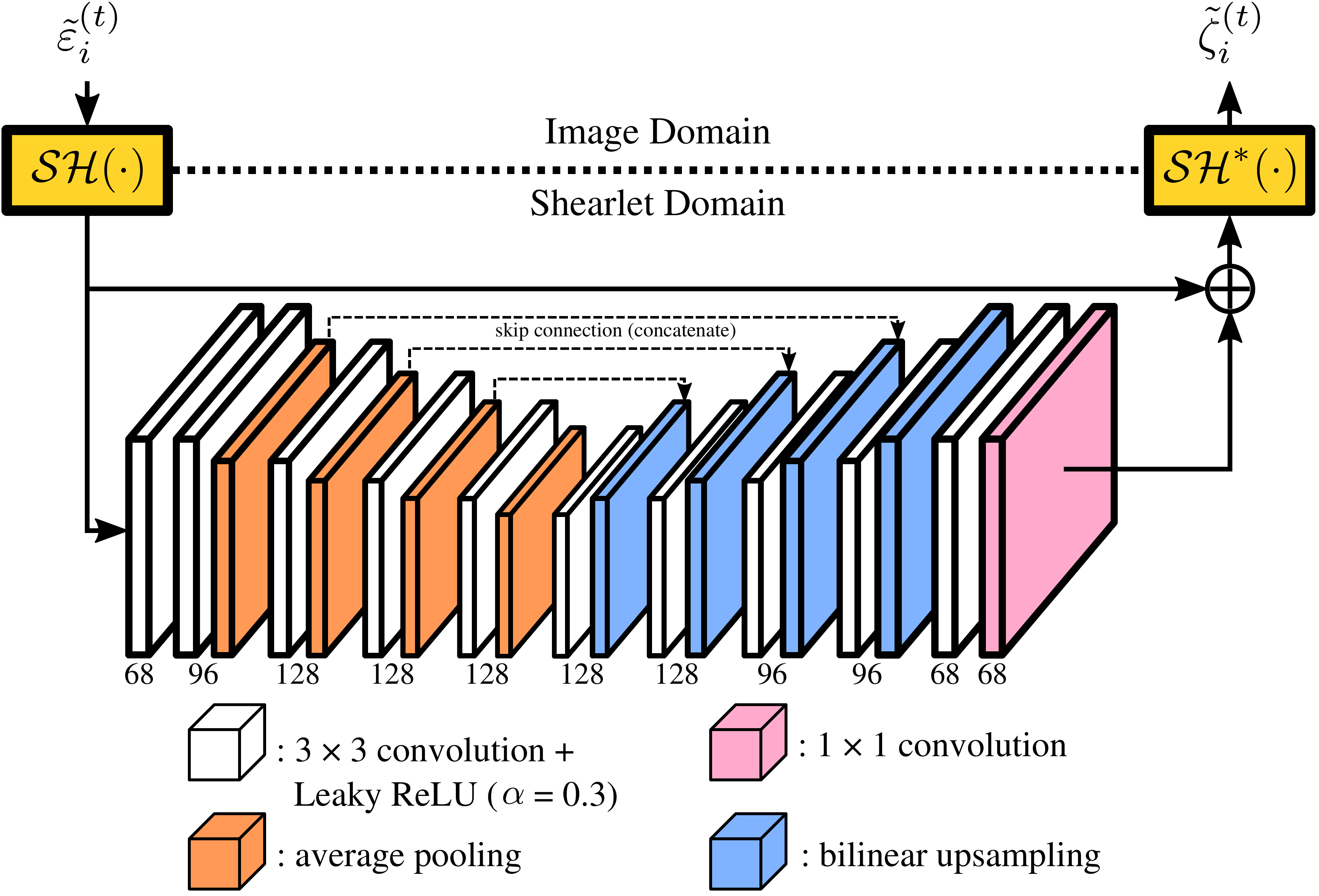}}
\end{minipage}
\vspace{-.8em}
\caption{Architecture of the deep convolutional network of CycleST, referred to as $\mathcal{R}(\cdot)$. 
Based on the U-Net architecture
\cite{ronneberger2015u} and residual learning strategy
\cite{he2016deep}, 
$\mathcal{R}(\cdot)$ refines the shearlet coefficients of $\tilde{\varepsilon}_i^{(t)}$ in shearlet domain to produce inpainted $\tilde{\zeta}_i^{(t)}$ in image domain.}
\vspace{.4em}
\label{fig:architecture}
\end{figure}

\subsubsection{Sparse regularization}
The remapped EPI $\tilde{\varepsilon}_i^{(t)}$ is then converted into shearlet coefficients via the shearlet analysis transform $\mathcal{SH}(\cdot)$. 
The sparse regularization step is essentially refining these coefficients in shearlet domain to fulfill image inpainting on $\tilde{\varepsilon}_i^{(t)}$ in image domain. 
To this end, the state-of-the-art deep learning techniques are exploited.
In particular, a deep network using cycle consistency loss with self-supervision setup is designed for the reconstruction of the shearlet coefficients.
The details of the network architecture and loss function of CycleST network are described as below.

\par
\noindent\textbf{\emph{Network architecture}}.
As shown in \figref{fig:architecture}, 
a residual convolutional network, 
based on the architectures of U-Net 
\cite{ronneberger2015u} 
and the generator network of CycleGAN 
\cite{zhu2017unpaired}, 
is adapted to perform the reconstruction of shearlet coefficients.  
The input data are $\tilde{\varepsilon}_i^{(t)}$, $t\in\{3, 5\}$, in image domain,
which can be converted into coefficients $\mathcal{SH}\Big(\tilde{\varepsilon}_i^{(t)}\Big)$ with $\eta=68$ channels in shearlet domain.
These coefficients are then fed to the encoder-decoder network of CycleST, represented by $\mathcal{R}(\cdot)$, to predict the residuals for the input coefficients, \ie $\mathcal{R}\Big(\mathcal{SH}\big(\tilde{\varepsilon}_i^{(t)}\big)\Big)$.  
The encoder component of $\mathcal{R}(\cdot)$ has four hierarchies, of which each is composed of one $3\times 3$ convolution, one Leaky ReLU and one average pooling layers. 
The decoder part also has four hierarchies. 
Each one consists of the same convolution and Leaky ReLU layers as that of the encoder, 
but a bilinear upsampling layer instead of the pooling layer. 
The skip connections concatenate the outputs of last three hierarchies of both encoder and decoder. 
It can also be seen that the last layer of $\mathcal{R}(\cdot)$ is only a $1\times 1$ convolution layer, without Leaky ReLU placed behind it.
Following the residual learning strategy
\cite{he2016deep}, 
the predicted coefficient residuals $\mathcal{R}\Big(\mathcal{SH}\big(\tilde{\varepsilon}_i^{(t)}\big)\Big)$ are merged with $\mathcal{SH}\Big(\tilde{\varepsilon}_i^{(t)}\Big)$ via an element-wise addition operation.
The output data of CycleST network are the corresponding densely-sampled EPIs $\tilde{\zeta}_i^{(t)}$, $t\in\{3, 5\}$, 
which can be written as below:
\begin{IEEEeqnarray}{c}	\label{eq:}
\tilde{\zeta}_i^{(t)} = \mathcal{SH}^*\bigg( \mathcal{SH}\Big(\tilde{\varepsilon}_i^{(t)}\Big) + \mathcal{R}\Big(\mathcal{SH}\big(\tilde{\varepsilon}_i^{(t)}\big)\Big) \bigg)\,.
\end{IEEEeqnarray}

\par
\noindent\textbf{\emph{Loss function}}.
Two kinds of losses are considered in the loss function of CycleST network, \ie 
the EPI reconstruction loss $\mathcal{L}_\mathrm{s}(t)$ and  cycle consistency loss $\mathcal{L}_{\cyc}$.
Both of them employ $\ell_1$ norm, 
since recent research indicates that $\ell_1$ norm is superior over $\ell_2$ norm for learning-based view synthesis and image inpainting tasks
\cite{niklaus2017iccv,niklaus2017cvpr,yu2019free}. 
The overall loss function $\mathcal{L}$ is a linear combination of $\mathcal{L}_\mathrm{s}(t)$, $t\in\{3, 5\}$, and $\mathcal{L}_{\cyc}$, \ie %that can be written as below:
\begin{IEEEeqnarray}{c}	\label{eq:loss_all}
\mathcal{L} =  \mathcal{L}_\mathrm{s}(3) + \mathcal{L}_\mathrm{s}(5) + \lambda\mathcal{L}_{\cyc}\,.
\end{IEEEeqnarray}
The EPI reconstruction loss $\mathcal{L}_\mathrm{s}(t)$ measures the reconstruction errors between ground-truth sparsely-sampled $\epsilon_i^{\crop}$ and predicted densely-sampled $\tilde{\zeta}^{(t)}$ in a self-supervised manner:
\begin{IEEEeqnarray}{c} \label{eq:loss_s}
\mathcal{L}_\mathrm{s}(t) =  \norm{\tilde{\zeta}_i^{(t)}\Big(1:\tfrac{(t-1)\tau}{n'-1} : \big((t-1)\tau  + 1\big) \Big) - \epsilon_i^{\crop}}_1.
\end{IEEEeqnarray}
The cycle consistency loss $\mathcal{L}_{\cyc}$ calculates the reconstruction differences between the predicted $\tilde{\zeta}^{(3)}$ and $\tilde{\zeta}^{(5)}$:
\begin{IEEEeqnarray}{c}	\label{eq:loss_cyc}
\mathcal{L}_{\cyc} =  \norm{\tilde{\zeta}_i^{(3)}\Big(1:(2\tau + 1) \Big) - \tilde{\zeta}_i^{(5)}\Big(1: 2: (4\tau+1)\Big) }_1 .
\end{IEEEeqnarray}
Finally, $\lambda$ is empirically set to 2.
\subsubsection{Post-shearing}
The post-shearing operation is only used in the inference phase of CycleST.  
In terms of using CycleST for reconstructing $\mathcal{D}$ from $\mathcal{S}$,
the input $\mathcal{S}$ has been sheared with parameter $\rho^{\mathcal{S}}$ in the pre-shearing stage. 
The post-shearing step compensates this through the same shearing operation on $\tilde{\zeta}_i^{(n)}$ with a new shearing parameter $-\frac{\rho^{\mathcal{S}} }{\tau}$. 
The sheared $\tilde{\zeta}_i^{(n)}$ is then cut by only keeping the top $\big((n-1)\tau  + 1\big)$ rows of it to produce $\zeta_i$ of the target $\mathcal{D}$.
It is suggested that $\rho^{\mathcal{S}}=d_{\min}^\mathcal{S}$.

\begin{table}[!t]
\begin{center}
\caption{Details about the training and evaluation datasets.} 
\label{tab:eva}
\vspace{-.8em}
\small
\scalebox{.85}{
\begin{tabular}{@{}lccccccc@{}}
%\hline
\specialrule{.15em}{.0em}{.2em}
3D light fields & $m$ & $l$ & $n'$ & $\ddot{n}$ & $\delta$ & n & $\dot{n}$  \\
%\hline
\midrule
$\mathcal{S}'_j$, $1 \leqslant j \leqslant 78$ & 512 & 512 & 9 & - & - &  - & - \\[.5ex]
$\Psi^1_j$, $1 \leqslant j \leqslant 9$ & 1280 & 720 & - & 97 & 16 & 7 & 193 \\[.5ex]
$\Psi^2_j$, $j\in\{1,2\}$ & 960 & 720 & - & 97 & 16 & 7 & 193 \\[.5ex]
$\Psi^2_j$, $j\in\{3,4\}$ & 960 & 720 & - & 97 & 32 & 4 & 97 \\ 
%\hline 
\specialrule{.15em}{.2em}{.0em}
\end{tabular}
}
\end{center}
%\vspace{-0.8em}
\end{table}

%$\epsilon_i^{\shear}=\shear(\varepsilon, d)$

\vspace{-.5em}
\section{Experiments}	\label{sec:expri}
%\vspace{-.5em}
\subsection{Experimental Settings}

\subsubsection{Training dataset}
The Inria synthetic light field datasets contains 39 synthetic 4D SSLFs with disparities from -4 to 4 pixels \cite{shi2019framework}, 
satisfying the requirement $d_{\range}^{\mathcal{S}'} \leqslant 8$ pixels in \secref{sec:symbols} for the training SSLF $\mathcal{S}'$. 
These 4D SSLFs have the same angular resolution $9\times 9$ and spatial resolution $512 \times 512$ pixels. 
%To avoid redundant training EPIs of the 4D SSLF data, 
We only pick the 5-th row and 5-th column 3D SSLFs from each synthetic 4D SSLF. 
Therefore, the training data of CycleST consists of $\mathcal{S}'_j$, $1 \leqslant j \leqslant 78$, $l=512$, $m= 512$ and $n'=9$.
The pre-shearing operation in \secref{sec:shear} is repeated three times for each $\mathcal{S}'_j$ with different shearing parameters $\rho^{\mathcal{S}'_j} \in \Big\{
\Big(d_{\min}^{\mathcal{S}'_j} - \big(\tau' - d_{\range}^{\mathcal{S}'_j}\big) \Big),
\Big(d_{\min}^{\mathcal{S}'_j} - 0.5\cdot\big(\tau' - d_{\range}^{\mathcal{S}'_j}\big) \Big),
 d_{\min}^{\mathcal{S}'_j} \Big\}$.
As a result, the number of the generated $\epsilon^{\crop}_i$ for each training epoch is $78\times 3 \times l = 119,808$.

\subsubsection{Evaluation Datasets}
Since there are few public real-world DSLF datasets, 
two evaluation datasets of light fields with \emph{tiny disparity ranges} ($\leqslant$ 2 pixels) are considered for the performance evaluation of different DSLF reconstruction methods. 
The Evaluation Dataset 1 (ED1) is the tailored High Density Camera Array (HDCA) dataset 
\cite{ziegler2017acquisition}
using the same cutting and scaling strategy as in  
\cite{gao2019tcsvt}.
Consequently, nine \emph{tiny-disparity-range} light fields $\Psi^1_j$, $1 \leqslant j \leqslant 9$, form ED1 with the same spatial resolution ($m\times l = 1280 \times 720$ pixels) and angular resolution ($\ddot{n}=97$).
%Note that all these nine light field scenes contain many repetitive-pattern objects and complex occlusions.  
For each $\Psi^1_j$, an input SSLF $\mathcal{S}^1_j$ can be produced using an interpolation rate $\delta = 16$. 
As a result, the angular resolution of $\mathcal{S}^1_j$ is $n=\big(\frac{\ddot{n}-1}{\delta} + 1\big) = 7$.  
The target $\mathcal{D}^1_j$ to be reconstructed from $\mathcal{S}^1_j$ has angular resolution $\dot{n}=193$.
%The $d_{\min}^{\mathcal{S}^1_j}$ and $d_{\range}^{\mathcal{S}^1_j}$ are shown in \tabref{tab:res1}.
%\subsubsection{Evaluation Dataset 2 (ED2)}
The MPI light field archive 
contains two \emph{tiny-disparity-range} light fields (`bikes' and `workshop') and two DSLFs (`mannequin' and `living room')
\cite{kiran2017towards},
which 
constitute the Evaluation Dataset 2 (ED2), \ie $\Psi^2_j$, $1 \leqslant j \leqslant 4$.
The spatial and angular resolutions of each $\Psi^2_j$ is $m=960$, $l=720$ and $\ddot{n}=97$. 
For the \emph{tiny-disparity-range} $\Psi^2_j$, $j\in\{1,2\}$, the interpolation rate $\delta$ is set to 16, such that $n=7$ and $\dot{n}=193$.
Regarding the densely-sampled $\Psi^2_j$, $j\in\{3,4\}$, the interpolation rate $\delta$ is set to the same as $\tau=32$, 
such that $n=4$ and $\dot{n}=97$.
The angular and spatial resolutions of the training and evaluation datasets are also summarized in \tabref{tab:eva}.
The minimum disparity and disparity range of $\mathcal{S}^1_j$ and $\mathcal{S}^2_j$ are exhibited in \tabref{tab:res1} and \tabref{tab:res2}, respectively.
%The four light field scenes here have no repetitive-pattern objects and less complex occlusions compared with ED1.  
%The $d_{\min}^{\mathcal{S}^2_j}$ and $d_{\range}^{\mathcal{S}^2_j}$ are depicted in \tabref{tab:res2}.

%\subsubsection{Evaluation Dataset 2 (ED2)}
%The five real-world DSLFs of the CIVIT DSLF dataset 
%\cite{civit}
%prepared for IEEE ICME 2018 grand challenge on DSLF reconstruction constitute the ED2, which can be represented by $\Psi^2_j$, $1 \leqslant j \leqslant 5$.
%The angular and spatial resolutions of each $\Psi^2_j$ are $m=1280$, $l=720$ and $\ddot{n}=193$.
%Since $\Psi^2_j$ is densely-sampled, the interpolation rate $\delta$ is set to the same as $\tau=32$ to generate the corresponding input $\mathcal{S}^2_j$ with angular resolution $n=7$.
%The target $\mathcal{D}^2_j$ to be reconstructed from $\mathcal{S}^2_j$ has angular resolution $\dot{n}=193$.

\subsubsection{Implementation details}
The weights of all the filters of the CycleST network $\mathcal{R}(\cdot)$ are initialized by means of the He normal initializer 
\cite{he2015delving}.
The AdaMax optimizer 
\cite{kingma2014adam}
is employed to train the model for 12 epochs on an Nvidia GeForce RTX 2080\,Ti GPU for around 33 hours. 
The learning rate is gradually reduced from $10^{-3}$ to $10^{-5}$ using an exponential decay schedule during the first four epochs and then fixed to $10^{-5}$ for the rest eight epochs. 
The mini-batch for each training step is composed of two different samples $\epsilon_i^{\crop}$ shown in \figref{fig:epi_all}\,(b).
The number of the trainable parameters of $\mathcal{R}(\cdot)$ is around 1.4\,M.
The implementation of ST is from \cite{gao2019light}.

\begin{table*}[t]
\begin{center}
\caption{Disparity estimation, minimum and average per-view PSNR results (in dB) for the performance evaluation of different light field reconstruction methods on ED1. 
The best two results are highlighted in \textcolor{red}{red} and \textcolor{blue}{blue} colors.
} 
\label{tab:res1}
\vspace{-.8em}
\scalebox{0.7}{
\begin{tabular}{@{}l|c|c|c|c|c|c|c|c|c|c@{}}
\specialrule{.2em}{.0em}{.3em}
%\toprule[\heavyrulewidth]
\multirow{2}{*}{$j$} & \multicolumn{2}{c|}{\textbf{Disparity (pix)} } & \multicolumn{8}{c}
{\textbf{Minimum PSNR / Average PSNR (dB)} } \\ 
\noalign{\vskip2pt}
\hhline{~|----------} 
\noalign{\vskip2pt}
%\cmidrule{2-3} \cmidrule{5-12} \\
  & $d_{\min}^{\mathcal{S}^1_j}$ & $d_{\range}^{\mathcal{S}^1_j}$ & SepConv ($\mathcal{L}_1$) \cite{niklaus2017iccv} & PIASC ($\mathcal{L}_1$) \cite{gao2018icmew} & DAIN \cite{bao2019depth}  & LVI \cite{xu2019quadratic} & QVI \cite{xu2019quadratic} & ST \cite{vagharshakyan2017accelerated} & DRST \cite{gao2019tcsvt} & CycleST
  \\ [.1ex] 
%\hline
\midrule
1 & 25 & 19   & 19.988 / 21.769 & 19.978 / 21.760 & 29.042 / 32.664 & 32.382 / 33.397 & 32.164 / 33.475 & 32.133 / \textcolor{blue}{35.185} & \textcolor{blue}{32.452} / 35.027 &  \textcolor{red}{34.288} / \textcolor{red}{35.918} \\
2 & 27 & 22   & 20.777 / 23.978 & 20.782 / 24.015 & 22.563 / 24.516 & 24.828 / 26.073 & 24.854 / 26.090 & \textcolor{red}{25.877} / \textcolor{red}{27.953} & 23.811 / 25.512 & \textcolor{blue}{25.409} / \textcolor{blue}{26.712} \\
3 & 28 & 27   & 24.081 / 26.969 & 24.089 / 27.013 & 25.077 / 27.794 & 27.940 / \textcolor{blue}{29.660} & \textcolor{blue}{28.292} / 29.724 & 26.672 / 29.403 & 26.725 / 28.622 & \textcolor{red}{28.614} / \textcolor{red}{30.841} \\
4 & 25 & 30   & 24.648 / 28.486 & 24.660 / 28.584 & 27.125 / 28.765 & 28.482 / 30.225 & 28.552 / 30.115 & 29.153 / \textcolor{red}{32.639} & \textcolor{blue}{29.162} / 31.179 & \textcolor{red}{29.320} / \textcolor{blue}{31.470} \\
5 & 25 & 29   & 26.942 / 29.060 & 26.954 / 29.135 & 28.330 / 29.739 & 30.129 / 31.095 & 30.361 / 31.173 & \textcolor{blue}{30.780} / \textcolor{red}{33.111} & 30.737 / 31.637 & \textcolor{red}{31.177} / \textcolor{blue}{31.913} \\
6 & 25 & 29   & 26.965 / 29.620 & 26.977 / 29.692 & 31.003 / 34.817 & 32.588 / 34.198 & 31.796 / 34.126 & 33.853 / 36.354 & \textcolor{blue}{34.118} / \textcolor{blue}{36.712} & \textcolor{red}{36.006} / \textcolor{red}{37.513} \\
7 & 26 & 17   & 21.223 / 24.750 & 21.224 / 24.784 & 22.645 / 24.718 & 24.488 / 26.202 & 24.760 / 26.252 & \textcolor{red}{25.458} / \textcolor{red}{27.876} & 24.458 / 26.423 & \textcolor{blue}{25.428} / \textcolor{blue}{27.024} \\
8 & 28 & 21   & 21.152 / 24.309 & 21.158 / 24.360 & 22.320 / 24.633 & 24.627 / 26.122 & 24.724 / 25.974 & \textcolor{blue}{26.137} / \textcolor{red}{28.451} & 24.500 / 26.549 & \textcolor{red}{26.301} / \textcolor{blue}{28.046} \\
9 & 28 & 27   & 26.455 / 29.750 & 26.468 / 29.839 & 26.791 / 30.658 & 30.451 / 31.636 & \textcolor{blue}{30.829} / 32.069 & 29.721 / \textcolor{blue}{32.252} & 29.169 / 31.513 & \textcolor{red}{31.745} / \textcolor{red}{33.963} \\
\specialrule{.2em}{.2em}{.0em}
\end{tabular}
}
\end{center}
\vspace{-.8em}
\end{table*}

\begin{table*}[!t]
\begin{center}
\caption{Disparity estimation, minimum and average per-view PSNR results (in dB) for the performance evaluation of different light field reconstruction methods on ED2. 
The best two results are highlighted in \textcolor{red}{red} and \textcolor{blue}{blue} colors.
} 
\label{tab:res2}
\vspace{-.8em}
\scalebox{0.7}{
\begin{tabular}{@{}c|c|c|c|c|c|c|c|c|c|c@{}}
\specialrule{.2em}{.0em}{.3em}
\multirow{2}{*}{$j$} & \multicolumn{2}{c|}{\textbf{Disparity (pix)} } & \multicolumn{8}{c}
{\textbf{Minimum PSNR / Average PSNR (dB)} } \\ [.5ex] 
\noalign{\vskip2pt}
\hhline{~|----------} 
\noalign{\vskip2pt}
  & $d_{\min}^{\mathcal{S}^2_j}$ & $d_{\range}^{\mathcal{S}^2_j}$  & SepConv ($\mathcal{L}_1$) \cite{niklaus2017iccv} & PIASC ($\mathcal{L}_1$) \cite{gao2018icmew} & DAIN \cite{bao2019depth}  & LVI \cite{xu2019quadratic} & QVI \cite{xu2019quadratic} & ST \cite{vagharshakyan2017accelerated} & DRST \cite{gao2019tcsvt} & CycleST
  \\ [.1ex] 
%  \hline
\midrule
1 & -14 &  23.5  & \textcolor{blue}{30.611} / \textcolor{blue}{32.994} & \textcolor{red}{30.845} / \textcolor{red}{33.012} & 29.625 / 31.032 & 29.449 / 31.470 & 29.752 / 31.548 & 29.932 / 32.804 & 29.775 / 31.712 & 29.845 / 31.645 \\
2 & -6.5 & 23   & 34.155 / 37.138 & \textcolor{blue}{34.324} / \textcolor{red}{37.363} & 33.186 / 34.341 & 34.013 / 35.254 & 34.300 / 35.410 & 33.911 / \textcolor{blue}{37.286} & 34.107 / 35.887 & \textcolor{red}{34.773} / 36.138 \\
3 & -15 & 29   & 31.571 / \textcolor{blue}{34.117} & \textcolor{blue}{31.662} / \textcolor{red}{34.290} & \textcolor{red}{31.789} / 32.964 & 30.806 / 32.710 & 31.071 / 33.008 & 30.849 / 33.610 & 31.513 / 33.775 & 31.453 / 33.615 \\
4 & -12 & 28  & 37.106 / \textcolor{blue}{41.760} & \textcolor{red}{37.371} / \textcolor{red}{42.797} & \textcolor{blue}{37.198} / 40.341 & 36.849 / 39.368 & 36.793 / 39.616 & 36.069 / 40.104 & 36.444 / 40.415 & 36.924 / 40.610 \\
%\hline 
\specialrule{.2em}{.2em}{.0em}
\end{tabular}
}
\end{center}
\vspace{-.8em}
\end{table*}

\vspace{-.5em}
\subsection{Results and analysis}	\label{sec:rea}
The proposed CycleST method is compared with the state-of-the-art video frame interpolation approaches,  \ie SepConv ($\mathcal{L}_1$) \cite{niklaus2017iccv}, DAIN \cite{bao2019depth}, LVI \cite{xu2019quadratic}, QVI \cite{xu2019quadratic}, and DSLF reconstruction methods, \ie PIASC ($\mathcal{L}_1$) \cite{gao2018icmew}, ST \cite{vagharshakyan2017accelerated}, DRST \cite{gao2019tcsvt}, on the above two evaluation datasets. 
The performance of all the algorithms is compared using their minimum and average per-view PSNR values on each evaluation light field.
The ED1 has nine evaluation light fields from nine different complex scenes containing many repetitive-pattern objects.  
The performance results on ED1 are exhibited in \tabref{tab:res1}.
It can be seen from this table that all the results of CycleST are either the best or the second best,
%CycleST performs the best on all the nine evaluation light fields except for $\Psi_2^1$ and $\Psi_7^1$.
%On $\Psi_7^1$, the minimum PSNR of CycleST is only 0.03 dB less than that of the best algorithm, \ie ST.
%Regarding the average PSNR, CycleST outperforms the others on $\Psi_j^1$, $j\in\{1,3,6,9\}$.
%; however, ST achieves the best performance on the other five evaluation light fields. 
suggesting that the proposed method can effectively handle the DSLF reconstruction on SSLFs with repetitive patterns and large disparity ranges.
%It can also be found that the performance of SepConv and PIASC are significantly worse than the other approaches on all the light fields in terms of both minimum and average PSNRs.
%There are two main reasons for this: 
%(i) both methods can hardly handle the interpolation for repetitive patterns, while all the nine evaluation light fields have repetitive-pattern objects; 
%(ii) the spatially adaptive kernels of both methods have a fixed size ($51 \times 51$),
%implying that the input SSLFs with large disparities are difficult for them to reconstruct.  
The ED2 has four evaluation light fields from four different scenes that are less complex than those of ED1.
Specifically, the scenes of ED2 have no repetitive-pattern objects and less occlusions compared with ED1. 
The DSLF reconstruction results of all the methods on ED2 are presented in \tabref{tab:res2}.
As can be seen from the table, CycleST achieves the best minimum PSNR result on $\Psi^2_2$. 
In addition, it can also be seen that all the results of PIASC are either the best or the second best; 
however, in \tabref{tab:res1}, the results of PIASC and SepConv on ED1 are significantly worse than the other baseline approaches. 
%achieve the best minimum and average PSNR results on the most of evaluation light fields of ED2, because each $\Psi^2_j$ has no repetitive patterns and $\disp\big(\mathcal{S}^2_j\big)$ can be well handled by the adaptive kernel of PIASC.
This implies that the kernel-based PIASC and SepConv are not as robust as CycleST for DSLF reconstruction,
since they may fail in DSLF reconstruction on large-disparity-range SSLFs with repetitive patterns and complex occlusions. 
Moreover, the proposed CycleST method outperforms the flow-based QVI and LVI on ED1 and ED2 \wrt minimum and average PSNRs.
Finally, since CycleST and DRST are developed based on ST, the computation time of all these three methods is compared in \tabref{tab:time}.
It can be seen from the table that CycleST is at least 8.9 times faster than ST and 2.1 times faster than DRST.
%Finally, the representative parts of the reconstruction results \wrt the 90th image of $\Psi_6^1$ and the 58th image of $\Psi_2^2$ are displayed in \figref{fig:fhg} and \figref{fig:mpi}, respectively. 

\begin{table}[!t]
\begin{center}
\caption{The average computation time and speedup over ST of reconstructing $\zeta_i$ from a sparsely-sampled $\varepsilon_i \in \mathbb{R}^{m \times n \times 3}$.} 
\label{tab:time}
\vspace{-.8em}
\small
\scalebox{.85}{
\begin{tabular}{@{}cc|c|c|c@{}}
%\hline
\specialrule{.15em}{.0em}{.2em}
 m & n & ST \cite{vagharshakyan2017accelerated} & DRST \cite{gao2019tcsvt} & CycleST  \\%[.2ex] 
%\hline
\midrule
 1280 & 7  & 3324\,ms (1.0\,x) & 742\,ms (4.5\,x) & 275\,ms (12.1\,x) \\ 
 960 & 7  & 2257\,ms (1.0\,x) & 598\,ms (3.8\,x) & 253\,ms (8.9\,x) \\
 960 & 4  & 1792\,ms (1.0\,x) & 365\,ms (4.9\,x) & 177\,ms (10.1\,x) \\
%\hline 
\specialrule{.15em}{.2em}{.0em}
\end{tabular}
}
\end{center}
\vspace{-0.8em}
\end{table}

\vspace{-.5em}
\section{Conclusion}
This letter has presented a novel self-supervised DSLF reconstruction method, CycleST, which refines the shearlet coefficients of the densely-sampled EPIs in shearlet domain to perform the inpainting of them in image domain. 
The proposed CycleST takes full advantage of the shearlet transform, encoder-decoder network with residual learning strategy and two types of loss functions, \ie the EPI reconstruction and cycle consistency losses.
Besides, CycleST is trained in a self-supervised fashion solely on synthetic SSLFs with small disparity ranges.
Experimental results on two real-world evaluation datasets demonstrate that CycleST is extremely effective for DSLF reconstruction on SSLFs with large disparity ranges (16\,-\,32 pixels), complex occlusions and repetitive patterns. 
Moreover, CycleST achieves $\sim$\,9x speedup over ST, at least.

%\begin{table}[htb]
%\setlength\tabcolsep{0pt}
%\begin{threeparttable}
%\caption{Tail Dependence.}
%\label{Tails}
%%\begin{adjustbox}{width=1\textwidth}  % don't use it! it is cause of your troubles ...
%\begin{tabular*}{\linewidth}{@{\extracolsep{\fill}}
%                             *{13}{c}}
%    \toprule
%\multicolumn{13}{l}{bla bla bla \dots}              \\
%    \midrule
%bla & bla & bla & bla & bla & bla & bla
%    & bla & bla & bla & bla & bla & bla             \\
%bla & bla & bla & bla & bla & bla & bla
%    & bla & bla & bla & bla & bla & bla\tnote{***}  \\
%    \midrule[\heavyrulewidth]
%\end{tabular*}
%\smallskip
%\begin{tablenotes}\footnotesize%\small
%\item[***] If $\lambda_{L}$ or $\lambda_{U} \geq \mathit{Pr}[1\%]$,
%\item[**]  If $\lambda_{L}$ or $\lambda_{U} \geq \mathit{Pr}[5\%]$.
%\end{tablenotes}
%%\end{adjustbox}
%\end{threeparttable}
%\end{table}

\balance
\bibliographystyle{IEEEtran}
\bibliography{ref}

\end{document}